Ten years of Nature Physics

# The monopole movement

*The monopole picture for spin ice offers a natural description for a confounding class of materials. A paper published in Nature Physics in 2009 applied it to study the dynamical properties of these systems — sparking intense experimental and theoretical efforts in the years that followed.*

Claudio Castelnovo

The rich and complex behaviour of strongly correlated many-body systems is often best interpreted in terms of effective degrees of freedom which abstract from the microscopic constituents. Once identified, these can provide a natural framework for understanding the principal characteristics of the specific system under investigation.

Spin ice is a case in point. Coined as an analogy with water ice [Harris1997], the term proved to be an appealing and intuitive means to understand the geometric origin of the low-temperature thermodynamic properties of rare-earth pyrochlore magnets such as $Ho_2Ti_2O_7$ and $Dy_2Ti_2O_7$, most notably their residual entropy [Ramirez1999]. However, the extent to which the analogy worked seemed to involve an almost miraculous degree of fine-tuning. The Holmuium and Dysprosium magnetic moments have non-collinear easy axes and long-range, anisotropic dipolar interactions [Siddharthan1999; denHertog2000] — a rather complex scenario as far as classical spin systems go. Indeed, it took almost a decade until the mechanism for this apparent fine-tuning was explained mathematically [Isakov2005].

Even so, it was not until much later that the 'natural' degrees of freedom for describing this behaviour were identified: at low temperature, spin ice can be viewed as a vacuum hosting emergent magnetic monopole excitations [Castelnovo2008]. This breakthrough made it possible to connect the physics of spin ice with models and techniques from seemingly unrelated areas of research, such as Coulomb liquids and random walks [Castelnovo2011]. It also set off a hunt to prove that emergent monopoles were real and not just convenient book-keeping concepts.

It is with this backdrop that, writing in Nature Physics in 2009, Jaubert and Holdsworth focused their attention on the dynamical properties of spin ice [Jaubert2009], which at the time were poorly understood. They incorporated Coulomb interactions between monopoles into a pre-existing theory of non-interacting dynamics [Ryzhkin2005]. They devised an efficient computational scheme that combines the energetics of dilute magnetic monopoles with the entropics of bulk spin ice, permitting detailed studies of remarkably large systems. They were thus able to show that magnetic relaxation measurements in $Dy_2Ti_2O_7$ [Snyder2004] can be described in terms of the diffusive motion of monopoles in the presence of long range Coulomb interactions and an underlying network of 'Dirac strings' [Castelnovo2008] filling the quasi-particle vacuum. Finally, they examined the behaviour of spin ice in the presence of a magnetic field and the monopole density gradient this produces near a surface boundary.

Through their numerical simulations and direct comparison with experiment, Jaubert and Holdsworth achieved two major results. Firstly, they showed that the long range Coulomb interaction between the monopoles plays an important role in achieving good agreement between theory and experiment that relate to the dynamical properties of these systems. Secondly, they showed that this agreement can be obtained by assuming that the spins flip according to a temperature-independent characteristic time scale. This promoted the minimal model of the energetics of monopoles in spin ice to one of its dynamics.

These results laid the foundations to further work examining the properties of spin ice away from equilibrium [Slobinski2010, Klemke2011, Matthews2012, Bovo2012, Erfanifam2014, Kolland2012]. In particular, the observation of a characteristic temperature-independent monopole hopping time scale was the stepping stone to further theoretical investigations [Castelnovo2010, Levis2012, Slobinski2010, Mostame2014], and to substantiate their potential relevance to experiments on real materials [Giblin2011, Paulsen2014, Jackson2014]. This led to the discovery of new and intriguing connections between the physics of spin ice and other areas of research, spanning reaction diffusion processes, dimer adsorption, and kinetically constrained models. Spin ice has now become a laboratory of choice for the study of tuneable, slow dynamics.

Jaubert and Holdsworth's work also represents the first study of the (static) monopole density profile close to the surface of a spin ice sample in presence of an applied field. Their numerical results were later followed by analytical work [Ryzhkin2011]. Combined with the role of magnetic impurities, this establishes an interesting parallel between the physics of monopoles in spin ice and that of charges in semiconductors, which is yet to be thoroughly investigated.

The results obtained by Jaubert and Holdsworth were not left unchallenged. Several groups worked towards new experimental magnetic susceptibility data to compare with numerical simulations, in addition to those already available in the literature [Snyder2004]. The old data were indeed confirmed, but longer relaxation times were also observed at lower temperatures [Quilliam2011, Matsuhira2011, Yaraskavitch2012, Revell2012, Takatsu2013], posing a new puzzle for the field. The more recent results revealed characteristic relaxation time scales that grew faster than those implied by the Coulomb liquid model with temperature-independent single spin-flip dynamics used in Jaubert and Holdsworth's simulations.

Possible explanations put forward to account for this include a direct temperature dependence of the single-ion spin flip time scale [Revell2012, Takatsu2013], and the presence of magnetic impurities [Revell2012]. Magnetic impurities in these materials and their relation to response and relaxation properties have only recently started to be investigated systematically [Sala2014]. It is too soon to tell which mechanism may be at the root of this increased slowing down, but the challenge is certain to stimulate new and exciting research on spin ice and related materials.

Spin ice has now become a point of reference for fractionalised topological spin liquid behaviour in three dimensions. Nevertheless, many open questions remain and some of the issues raised by Jaubert and Holdsworth are still not entirely resolved.

A significant challenge that lies ahead is to understand how we can best use our knowledge of classical spin ice to study related quantum mechanical systems [Gingras2014]: What does a quantum Coulomb liquid look like and how can we detect it experimentally? How does the nature of its elementary fractional excitations affect its relaxation and response properties? Once again, an effective description in terms of emergent degrees of freedom will play a central role in modelling and understanding these quantum mechanical systems. The journey on this front has just begun.

Claudio Castelnovo works in the Theory of Condensed Matter Group of the Cavendish Laboratory at the University of Cambridge, 19 JJ Thomson Avenue, Cambridge CB3 0HE, UK.